# Predicting Generalization of AI Colonoscopy Models to Unseen Data

Joel Shor*[1], Carson McNeil*[1], Yotam Intrator[1], Joseph R Ledsam[2], Hiro-o Yamano[4], Daisuke Tsurumaru[3], Hiroki Kayama[2], Atsushi Hamabe[5], Koji Ando[3], Mitsuhiko Ota[3], Haruei Ogino[3], Hiroshi Nakase[2], Kaho Kobayashi[2], Masaaki Miyo[5], Eiji Oki**[3], Ichiro Takemasa**[5], Ehud Rivlin**[1], Roman Goldenberg**[1]

1 = Verily Life Sciences
2 = Google Japan
3 = Department of Surgery and Science, Graduate School of Medical Sciences, Kyushu University
4 = Sapporo Medical University School of Medicine, Department of Gastroenterology and Hepatology
5 = Department of Surgery, Surgical Oncology and Science, Sapporo Medical University
*authors share first authorship position
**authors share last authorship position

# Introduction

Early detection of colon cancer is critical for improving patient outcomes, yet the effectiveness of artificial intelligence (AI) models in colonoscopy is hampered by a model's ability to generalize to new patient populations. This limitation arises due to differences in patient age, demographics, colonoscopy procedures, and imaging technologies. This limitation also arises when applying a Computer Aided Detection (CADe) [1-3] model to countries that might have different healthcare systems than the training data [4-6]. These differences can potentially lead to inaccurate detection and compromising patient care [7-9], and the performance degradation can be extreme [10, 11]. Furthermore, ensuring performance parity across diverse subgroups is essential for equitable healthcare delivery [12-15].

Currently, evaluating AI model generalizability relies heavily on expensive and time-consuming labeled data. While some label-free methods exist, they are often domain specific [16-18]. In this work, we use self-supervised representation learning to create a label-free method of determining which data are most likely to cause an AI model to perform poorly. We demonstrate the effectiveness of this method using the current state-of-the-art colonoscopy CADe model [19] trained on Israeli data. We show that our method can predict which Japanese colonoscopy videos are most dissimilar to the training data, and therefore exhibit decreased detector performance. This information can help build models that generalize across populations and healthcare settings.

# Materials and Methods

## Computed Aided Detection (CADe) polyp detection task

Colorectal cancer (CRC) is the second leading cause of cancer death globally [5], and results in an estimated 900,000 deaths each year [20]. One primary tool for identifying CRC are screening colonoscopies where a clinical operator can identify and remove *polyps*, abnormal clumps of cells, which over time can become malignant [21]. Computer Aided Detection of polyps (CADe) uses video and image-based artificial intelligence (AI) systems to assist operators in detecting polyps. These systems are already used in prospective trials and clinics [2, 3].

The problem is usually formulated as object detection to predict bounding boxes that overlap with the polyps in the colonoscopy video in real time. In *Supplementary Section: Data: Colonoscopy examples*, we give examples of colonoscopy videos, annotations, and predictions made by a production-grade polyp detection system described in [19]. We describe the colonoscopy data used to train and evaluate this system in *Section: Data Collection and patients*, and a summary of the CADe system training in *Supplementary Section: CADe Architecture and Training*.

## Data collection and patients

**Israeli data collection**: 7,877 laparoscopy videos and 13,979 colonoscopy videos were collected from hospitals in Israel. Each case consisted of a single video of an entire procedure, using a variety of endoscope models. All videos and metadata were deidentified, according to the Health Insurance Portability and Accountability Act Safe Harbor.

**Japanese data collection**: This retrospective study was approved by the Ethics Committees of the Institutional Review Boards of Kyushu and Sapporo University Hospitals (Sapporo IRB: 322-157 Kyushu IRB: 2021-448). Colonoscopy videos were collected from two large university hospitals in Kyushu (86 videos) and Sapporo (268 videos), Japan. Videos were recorded at various resolutions, formats, and containers. All data were de-identified.

Videos were compressed using H264 at compression rate QP20, which has been shown to have minimal impact on polyp detection performance [22, 23]. For more details on data storage and preprocessing, see *Supplementary Section: Data Storage and Preprocessing.* For details on the polyp annotation procedure, see *Supplementary Section: Data Annotation*.

There are numerous differences between the colonoscopy videos, such as scoping hardware, endoscope software, screening guidelines, bowel preparation, and patient demographics [24, 25] (see *Supplementary Section: Data descriptives and differences between Israel and Japan*). Unlike the Israeli dataset, parts of the Japanese videos contain narrow-band imaging (NBI), a technology that modifies the wavelength and bandwidth of a colonoscope's light to better visualize the colonic mucosa, and chromoendoscopy (CE), a technique involving spraying dye to highlight edge contrasts [26]. Examples of these differences can be seen in Figure 1.

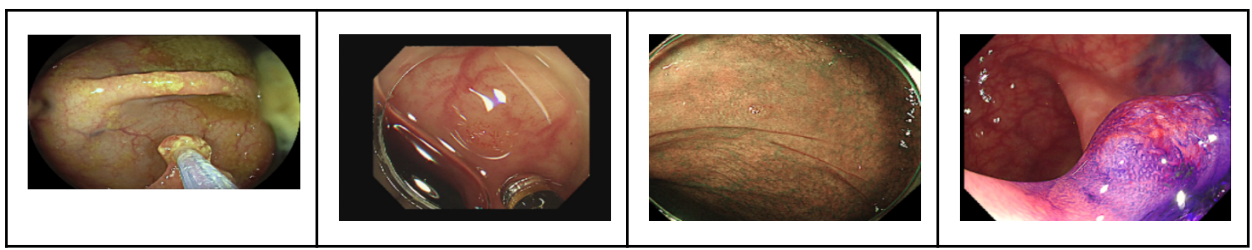

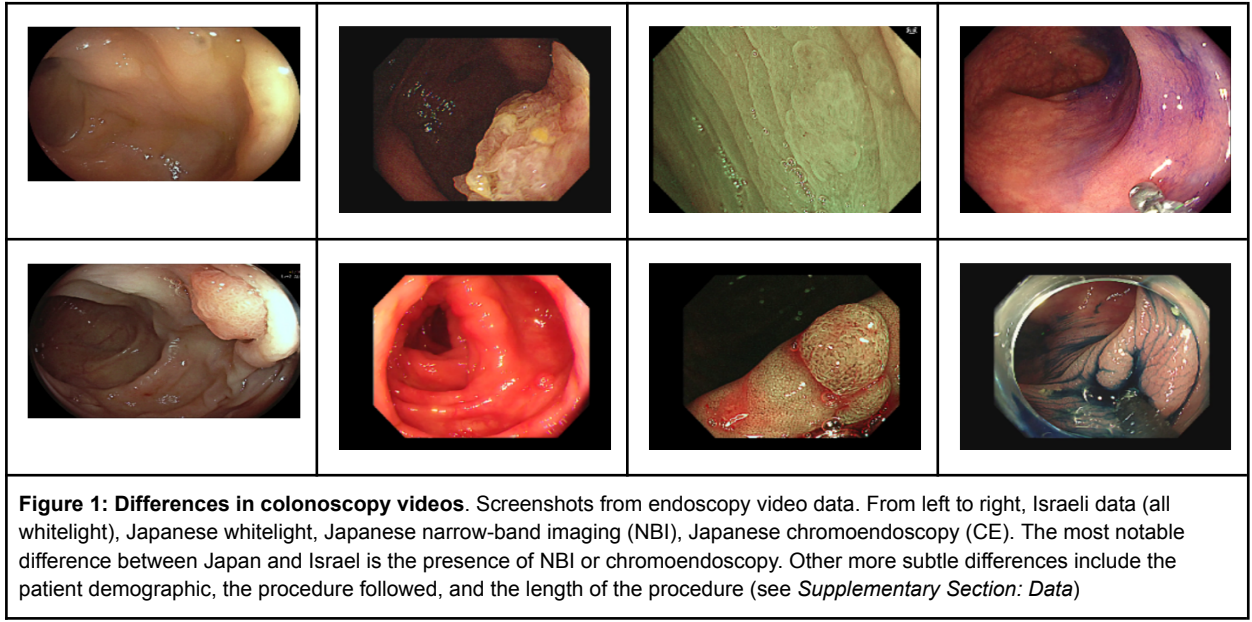

**Figure 1: Differences in colonoscopy videos**. Screenshots from endoscopy video data. From left to right, Israeli data (all whitelight), Japanese whitelight, Japanese narrow-band imaging (NBI), Japanese chromoendoscopy (CE). The most notable difference between Japan and Israel is the presence of NBI or chromoendoscopy. Other more subtle differences include the patient demographic, the procedure followed, and the length of the procedure (see *Supplementary Section: Data*)

## Masked Siamese Networks (MSN): Compactly representing medically-relevant colonoscopy information

Applying a CADe system to a new country requires some combination of recollecting data, re-annotating, and retraining, which can be time consuming and expensive. We present a technique for making this process cheaper and more efficient by quantifying differences in colonoscopy videos without the need for manual annotations. *We apply our technique to colonoscopy videos, but the technique is general enough to be applicable to other medical domains.*

We use embeddings described in [27]. These embeddings are trained using the Masked Siamese Networks (MSN) described in [28]. Importantly, MSN are trained in a self-supervised way (**no annotations required**) by trying to predict "masked out" sections of the image during training [29]. Representations learned from this proxy task and others have been used in a number of problems and modalities [30, 31]. Figure 2 provides a schematic explanation of how MSN is trained, and more details are provided in *Supplementary section: Learned Representations*. We evaluate MSN embeddings against other types of embeddings such as the SimCLR [32] model described in [33] and pretrained image embeddings, which we describe in *Supplementary section: Learned representations*.

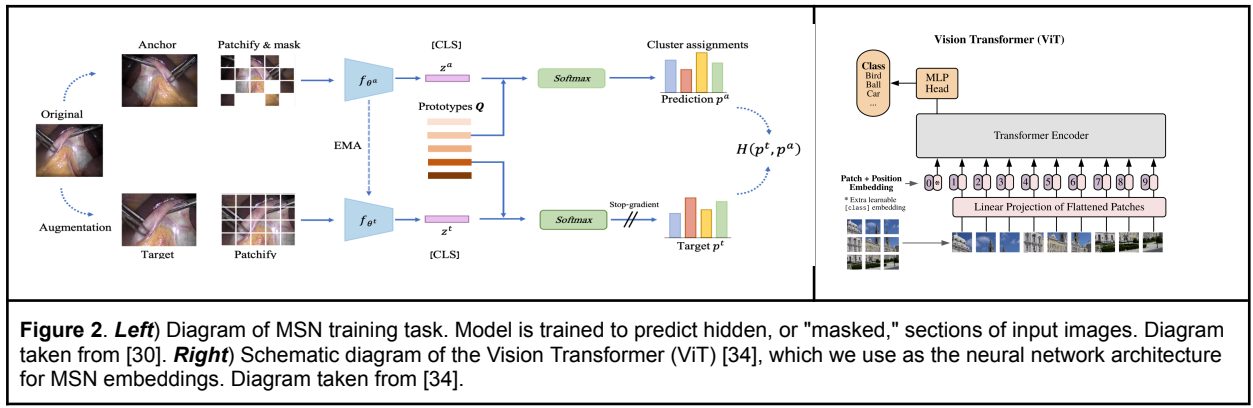

**Figure 2**. *Left*) Diagram of MSN training task. Model is trained to predict hidden, or "masked," sections of input images. Diagram taken from [30]. *Right*) Schematic diagram of the Vision Transformer (ViT) [34], which we use as the neural network architecture for MSN embeddings. Diagram taken from [34].

# Evaluation

MSN was trained to predict masked-out regions in whitelight, Israeli colonoscopy data (see section Materials and Methods for more details), which does not contain NBI or CE at all, differs from Japanese data in demographics of the patient population, endoscope hardware, and scoping procedure (see *Supplementary Section: Data descriptives and differences between Israel and Japan* for more details), and does not contain any direct information about the polyp detection task. We evaluate MSN on multiple tasks, and surprisingly find that MSN contains relevant information for all tasks, despite not having any Japanese or polyp detector information available during training:

1. Predicting the presence of NBI and CE in Japanese colonoscopy data
2. Predicting how well a polyp detector will perform on a) Israel (in-domain) and b) Japanese (out-of-domain) colonoscopy data

We compare MSN to several different baselines, including alternate self-supervised representations trained for polyp re-identification [31], off-the-shelf image representations from ImageNet Inception v3 [35, 36], and the features extracted from a randomly initialized neural network, referred to as a "Deep Image Prior" [37].

## Predicting NBI and CE in Japanese colonoscopies

### Unsupervised prediction of out-of-domain colonoscopy features

We use clustering techniques to show that MSN contains information about NBI and CE, despite not having seen either during training. We perform t-distributed stochastic neighbor embedding (t-SNE) [38], an unsupervised nonlinear clustering technique, and explore the substructure of the clustering. We quantify this exploration with bootstrapped confidence intervals of the Frechet distance between clusters (*Results Section: Determining Substructure (NBI/CE) without labels*). We use the Frechet distance instead of alternatives like Centered Kernel Alignment [39], which has also been used to compare embeddings [40], since it has strong theoretical foundations [41-43], and has many recent successful applications [44-45]. The Frechet distance is defined and explained in *Supplemental Section Evaluation details Frechet Distance*.

### Prediction of out-of-domain colonoscopy features

MSN embeddings can be used to train a classifier both on frames for which the ground truth label is accurate, such as NBI, or those with noisy labels, such as CE (see Section Differences in Japanese data). We train a SVM classifier using an RBF kernel [46, 47] using noisy labels, and compare the performance of the trained model and heuristic on a set of high-quality, manually annotated frames. In all cases, we split data randomly between train and test, and all evaluations are reported on the previously unseen test set (see Supplementary Methods).

### Predicting detector performance

Using the polyp detector described in *Supplementary Section: CADe Architecture and Training* and the polyp annotation procedure described in *Supplementary Section: Data Annotation*, we evaluate MSN's ability to predict detector performance on Israeli and Japanese colonoscopies. We use this embedding to predict the detector performance. To produce MSN embeddings, we crop the bounding box produced by the annotators or the detection model, and run inference on the ViT model trained according to *Supplementary Section: Learned representations: MSN architecture and training details*.

Video CADe models rely on a *threshold* parameter [19] to balance between false positives (incorrect detections) and false negatives (missed polyps). Detections with confidence scores above the threshold are considered real polyps, so high confidence scores on real polyp frames mean that the polyp will be correctly detected over a wider range of operating parameters, and lead to better detector performances. The same is true for low confidence scores on false positives. We use the confidence score as the prediction target. For more details on the exact prediction target, see *Supplementary Section: Detector performance target.*

***We evaluate MSN's ability to predict detector performance in three settings***: without any detector performance data (**unsupervised**), with detector performance only on Israeli colonoscopy data (**supervised, out-of-domain data**), and with a small number of detector performance on Japanese colonoscopies (**supervised, in-domain data**). We evaluate the detector on both Israeli colonoscopy data (in-domain for the detector and MSN) and Japanese colonoscopy data (out of domain for the detector and MSN).

# Results

## Detecting unseen techniques in out-of-domain data

### Determining substructure (NBI/CE) without labels

We qualitatively investigate MSN's ability to detect meaningful differences in data without labels by visualizing 384-d MSN embeddings in 2D using t-SNE [40] (see Figure 3). We find that anomalous WL points near the CE cluster (cluster 1) in Figure 3 revealed mislabelled CE polyps. We manually inspected all 142 frames in this cluster, and found that the accuracy of CE labels in cluster 1 from our CE heuristic (*Supplementary Differences in Japanese data: narrow-band imaging (NBI) and chromoendoscopy (CE)* ) was 0.87. If we instead label the entire cluster as "CE", as MSN geometry suggests, the true accuracy improves to 0.93 (not 1.0, due to the introduction of false positives). A similar type of qualitative procedure is possible to some degree using SimCLR embedding space, but not other controls (*Supplementary Figures S5, S6, S7*). Note that MSN was trained without any NBI or CE data, but can still accurately cluster them. For t-SNE on other embeddings, see *Supplementary Section: t-SNE Inspection of Other Embedding Spaces*.

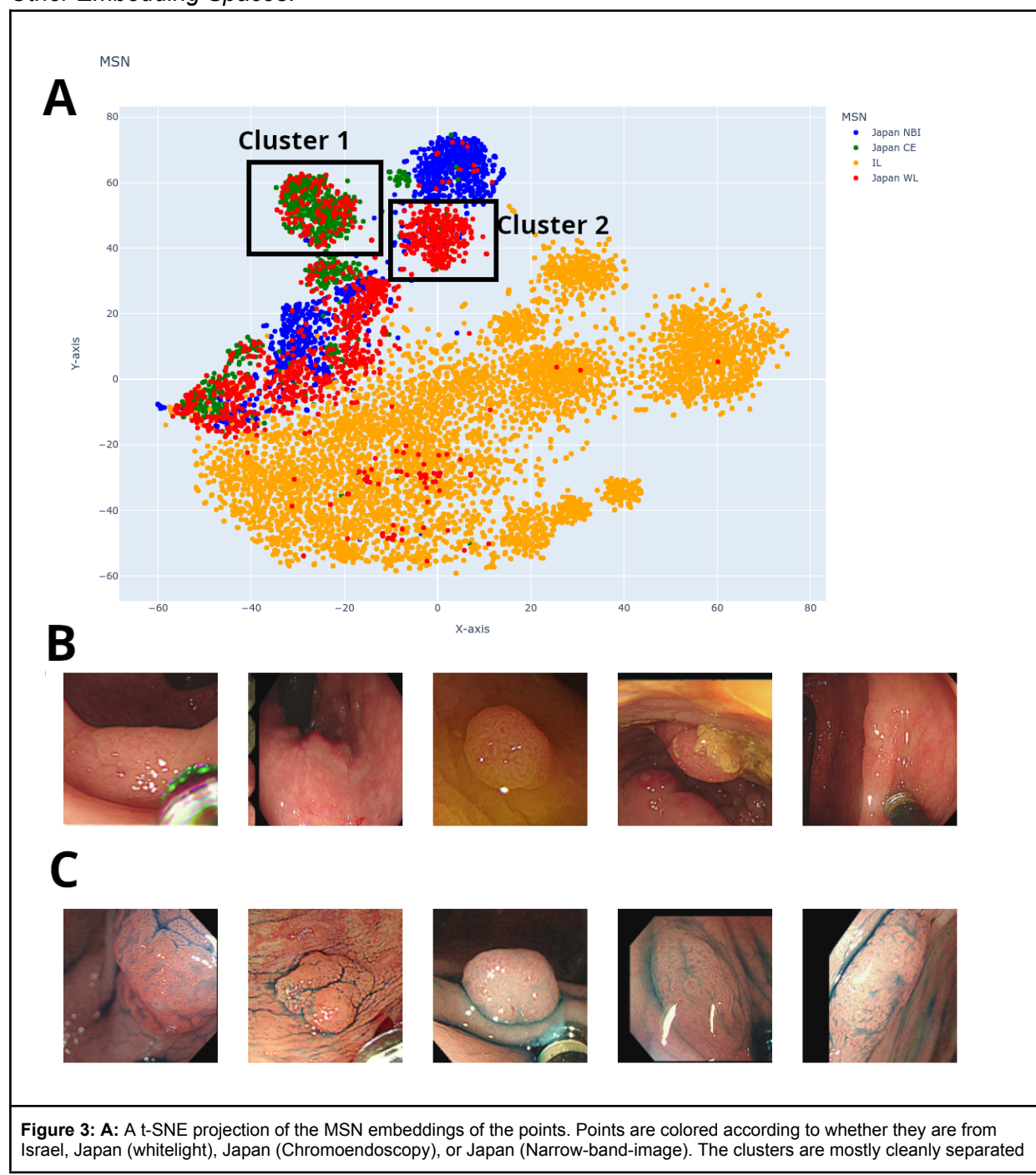

**Figure 3: A:** A t-SNE projection of the MSN embeddings of the points. Points are colored according to whether they are from Israel, Japan (whitelight), Japan (Chromoendoscopy), or Japan (Narrow-band-image). The clusters are mostly cleanly separated

to visual inspection, but the largest discrepancy can be seen by the large number of red (Japan WL) points that are in Cluster 2 instead of Cluster 1. **B** Samples Japanese WL points lying within Cluster 1, whereas **C** samples Japanese WL points lying within Cluster 2. We can see that those within Cluster 2 are in fact Chromoendoscopy images that were incorrectly labeled as WL. This demonstrates the utility of visually inspecting t-SNE spaces in investigations.

We quantify MSN's ability to detect meaningful differences in data without labels using the Frechet distance on MSN representations (see *Materials and Methods Section: Unsupervised prediction of out-of-domain colonoscopy features*). Japanese NBI and CE frames are consistently more dissimilar to the Israeli training data than Japanese white light using bootstrapped, two-sided difference of means z-test (t = -675.9, $p < 10^{-8}$ for NBI, z = -496.5, $p < 10^{-8}$ for CE). MSN is sufficiently discriminative for Israel, Japan whitelight, Japan NBI, and Japan CE are sufficiently different that 95% confidence intervals have no overlap. See Supplementary Table S2 for confidence intervals for the differences between these groups, for all embeddings.

## Determining substructure (NBI/CE) with few labels

We investigate MSN's ability to detect meaningful differences in data with labels by training a shallow model (SVM with RBF kernel) on MSN embeddings, and attempting to classify NBI vs not NBI and CE vs not CE. Neither NBI nor CE were included in MSN training data. After training on an 80:20 split (3571 frames in training), we find that shallow models are able to detect these phenomena (Table 3, top 2 rows). As noted above, the CE heuristic has label inaccuracies, so we also report performance on a manually annotated subset of 716 frames. The NBI labels came from the endoscope system and are perfect.

| Detection task | *MSN* | *SimCLR* | *Inception* | *Deep Image Prior* | *Heuristic* |
|---|---|---|---|---|---|
| NBI vs not NBI | **0.96 (0.95, 0.97)** | **0.97 (0.95, 0.98)** | 0.71 (0.64, 0.77) | 0.69 (0.63, 0.77) | N/A |
| CE vs not CE (manual labels) | **0.90 (0.88, 0.92)** | 0.88 (0.86, 0.90) | 0.61 (0.55, 0.68) | 0.62 (0.56, 0.69) | 0.79 (0.76, 0.82) |
| CE vs WL (heuristic labels) | **0.87 (0.84, 0.89)** | **0.87 (0.86, 0.90)** | 0.71 (0.64, 0.78) | 0.61 (0.54, 0.69) | N/A |
| CE vs WL (manual labels) | **0.90 (0.88, 0.93)** | 0.85 (0.82, 0.88) | 0.56 (0.47, 0.64) | 0.51 (0.43, 0.60) | 0.79 (0.76, 0.82) |

**Table 3:** Accuracy of SVM classifiers trained to predict various polyp features from MSN embeddings. Chromoendoscopy labels were manually corrected. We can see that there is significant disagreement between uncorrected CE labels, corrected CE labels, and the classifier. However the classifier matches manual labels closer. It can also be seen that while SimCLR too has this helpful effect, it overfits the noisy data more, at the cost of a lower true accuracy.

## Determining substructure (NBI/CE) with noisy labels

The heuristic CE labels have some amount of noise. We demonstrate that MSN can improve the quality of these noisy labels without any manual correction. We manually annotate a test split of 716 examples in order to evaluate our performance. Agreement between heuristic and manual annotation on the test set is .79 (bootstrapped 95% CI is [.76, .82]). We train an SVM with RBF kernel **on the noisy labels**, and find that the model outperforms the heuristic that generated the noisy labels (90% accuracy for MSN vs 79% for the heuristic). ***All of the other representations had worse performances on the manual annotations*** (Table 3, borrow 2 rows).

## Predicting CADe Polyp detector performance

### Predicting performance with few labels

Table 5 displays the performance of a supervised RBF-SVM trained to predict NLL confidence scores from MSN embeddings. By training only on detector confidences in Israel, we can achieve significantly-correlated predictions of detector confidences both in Israel and Japan. Including some labels from Japan improves this performance. MSN greatly outperforms other representations for all of these tasks. This approach is motivated by *Supplementary Figure S4*, which shows that t-SNE MSN clusters show some patterns with respect to detector performance.

| Train set | | Test set | | Embedding | | | |
|---|---|---|---|---|---|---|---|
| **Israel** | **Japan** | **Israel** | **Japan** | MSN | SimCLR | Inception | Deep Image Prior |
| ✅ | | ✅ | | **0.79 (0.77, 0.81)** | **0.78 (0.76, 0.80)** | 0.52 (0.49, 0.56) | 0.04 (-0.01, 0.10) |
| ✅ | | | ✅ | **0.37 (0.30, 0.43)** | 0.18 (0.11, 0.25) | 0.11 (0.04, 0.18) | 0.08 (0.01, 0.16) |
| ✅ | ✅ | | ✅ | **0.56 (0.44, 0.65)** | 0.34 (0.19, 0.48) | 0.11 (-0.05, 0.27) | 0.01 (-0.16, 0.16) |

**Table 5:** Resulting performance for training simple SVMs trained to predict polyp detector confidence from various embedding spaces. Performance summarized as a Pearson correlation between the predictions and actual detector confidence on the held-out TEST set. 95% confidence interval included.

# Discussion

This work proposes a label-free method of learning medical image representations, based on training a model to be impervious to masking out part of the image (MSN). When trained on pre-cropped polyp images in whitelight from Israeli colonoscopies, we demonstrate some useful and surprising properties:

1. We show that Euclidean and distributional distances between MSN representations correspond to human intuition of distance e.g. NBI is farther from Israeli whitelight than Japanese whitelight. MSN distances are meaningful in the complete absence of labels, on data from a country not seen during training (Japan), and on unseen techniques (NBI and CE).
2. MSN embeddings can be used to detect NBI and CE frames, both in an unsupervised way (t-SNE clustering) and with labels (classification with SVMs).

3. MSN is robust to label noise in a way that other representations are not.
4. Despite not being trained with any polyp detection labels or any data from Japan, MSN can predict the performance of polyp detectors on Israeli and Japanese colonoscopies. When given small amounts of in-domain detection-specific information, the predictions improve.

**The image-masking proxy task learns about polyp detection**. We show that the self-supervised masked-out proxy task can predict polyp detection performance in an unseen country. This result raises the question of what the properties of polyp detection are that allow it to be predicted from MSN embeddings. There are several possible explanations. First, the masking proxy task may be powerful enough to learn a compact representation of polyp images that captures information about many downstream tasks, including polyp detection but perhaps also other tasks. Second, MSN distances could correspond simply to "in distribution", and function essentially as an outlier detection.

**The image-masking proxy task on whitelight learns about NBI and CE.** We show that MSN can be trained on Israeli, whitelight-only data, and meaningfully generalize to NBI and CE in Japan. This could be due to MSN acting as an out-of-distribution detector, or because NBI and CE exist on some continuum that are found in whitelight data such as polyps with more defined edges or polyps with odd coloration.

**MSN alone can denoise CE detection.** Surprisingly, cluster assignment via t-SNE inspection of MSN space outperformed our heuristic in labeling CE. This substantial improvement of noisy labels is a low-supervision process enabled by using the MSN space as an intermediate tool. We also found that an SVM trained on MSN (without corrective labeling) could *automatically* denoise heuristic CE labels, thus achieving an improvement without any human input. This finding both provides immediate practical utility, and demonstrates that MSN derives a semantically-relevant representation, even for a modality on which it was never trained.

Our approach makes it possible to use label-free methods to quantify differences in data that might present challenges to model generalizability. Researchers can use our approach to preferentially collect more of the "difficult" data in order to improve model performance and to explore model performance ahead of clinical studies. Doing so may reduce the risk of degradation due to differences in data distribution.

This study has a number of limitations. First, future work should demonstrate the approach on a greater number of countries and sites, since practice can even vary between nearby sites [48], and. Second, future work can explore other colonoscopy tasks. Third, future work can explore whether the same findings hold in other medical imaging domains e.g. radiology. Fourth, future work can explore other factors that affect generalizability like variation in user ability [49] and differences in operational environment [50].

# Conclusion

We demonstrate that MSN can identify distribution shifts in clinical data and can predict CADe detector performance on unseen data, without labels. Our self-supervised approach can aid in detecting when data in practice is different from training, such as between hospitals or data has meaningfully shifted from training. MSN has potential for application to medical image domains beyond colonoscopy.